\documentclass[aps,pre]{revtex4}
\usepackage{amssymb}
\usepackage{amsmath}
\usepackage{graphicx}
\usepackage{float}
\usepackage{xcolor}
\begin{document}

\title{Non-linear oscillators with Kuramoto-like local coupling: Complexity analysis and spatiotemporal pattern generation.}
\author{K. \surname{Garc\'ia Medina}}
\affiliation{Facultad de F\'isica, Universidad de La Habana, San Lazaro y L. CP 10400. La Habana. Cuba.}

\author{E. \surname{Estevez-Rams}}
\email{estevez@fisica.uh.cu}
\affiliation{Facultad de F\'isica-Instituto de Ciencias y Tecnolog\'ia de Materiales(IMRE), Universidad de La Habana, San Lázaro y L. CP 10400. La Habana. Cuba.}

\author{D. \surname{Kunka}}
\affiliation{Institute of Microstructure Technology (IMT), Karlsruhe Institute of Technology (KIT), Hermann-von-Helmholtz-Platz 1, 76344 Eggenstein-Leopoldshafen, Germany}

\begin{abstract}
Can a simple oscillator system, as in cellular automata, sustain complex nature upon discretization in time and space? The answer is by no means trivial as even the most simple, two-state, nearest neighbours cellular automata can lead to Universal Turing Machine (UTM) computing power. This study analyses a recently proposed model consisting of a ring of identical excitable Adler-type oscillators with local Kuramoto-like coupling in terms of its complexity. Regions with non-trivial, complex behaviour have been identified, where spatiotemporal maps closely resemble those found in elementary cellular automata, not only from the visual perspective but also from entropic measures characterization. Also, the possibility of enhanced computation at the edge of chaos is explored by monitoring the effective complexity measure, entropy density and informational distance, following previous approaches. Distance matrix, and the corresponding dendrograms, show that in the complex regions, a non-trivial set of hierarchies emerges where local communities can appear and sustain themself in time. The fact that coupled oscillators can be realized through dedicated electronic circuitry or the result of natural processes can make finding such complex dynamics, with potential computational power, an important one in a broad scope of applications and implications.
\end{abstract}


\date{\today}
\maketitle

   
\section{Introduction}\label{Introduction}
Cellular Automata (CA) has been known to be simple systems exhibiting surprising behaviours, including chaotic and complex regimes \cite{wolfram86}. They are discrete time and space systems where the state of each unit depends on that of its nearest neighbours in the previous time step. It is even known that in the most simple type of CA, those with only two possible states $(0,1)$ and nearest-neighbour interaction, some dynamic rules can lead to Universal Turing Machine (UTM) \cite{wolfram02}. 

Consider now a system of coupled oscillators where each oscillator is simple and coupled through the same nearest-neighbour topology. Is it possible to find the same wealth of behaviours as those in cellular automata upon time and space discretization? The question is relevant as coupled oscillators can be realized through dedicated electronic circuitry or the result of natural processes \cite{Strogatz1993,Strogatz2001,Mosekilde,Zillmer}. To answer this question, we have been studying coupled oscillators, where each unit in isolation has a well-studied dynamics, even analytically solvable, and the coupling between them is kept as simple as possible \cite{Estevez2018}. Indeed, several regions of different dynamical natures have been identified in these systems, where a chaotic and complex region could be readily seen. Furthermore, the complex region showed interesting features in the spatiotemporal map, with long time and spatial correlations as the individual units couple themself in non-trivial manners \cite{Estevez2023}. Furthermore, evidence of what could be enhanced computation at the edge of chaos (EOC) was identified, which is a hallmark of complexity and has also been seen in CA when transitioning towards, or from, rules that behave as UTM \cite{Estevez2019}.

In the oscillator model previously studied, which we call the Alonso model, the coupling between units is realized according to the absolute value of the phase. However, a more common type of coupling will be where the relevant variable is the difference of phase between neighbouring units, the so-called Kuramoto coupling \cite{kuramoto75,acebron05}. Recently, this new model has been studied \cite{garcia22}. The first report was focused on understanding the possibility of synchronization processes among the oscillating units and their dynamic behaviour. No attempt was made to study complexity and chaos, which is the purpose of this contribution. We will be closely looking at the possible emergence of non-trivial correlations between chaotic and predictable behaviour, the fingerprint of complexity.

When the coupling among oscillators is weak, amplitudes evolution remains practically the same, and phase equations capture the dynamical features of the system \cite{Winfree1967}. One of the main advantages of using phase oscillators to study populations of interacting non-linear oscillators is that the equations are more amenable to analytic approaches while describing the relevant dynamical features of the system.  
  
One of the most representative examples of this approach is the model proposed by Kuramoto \cite{kuramoto75,acebron05} when studying synchronization processes in large populations of almost identical weakly coupled non-linear oscillators. There is no straightforward approach to the general problem described by Kuramoto since it strongly depends on the particular topology and features of interactions in the system. Kuramoto's original model focused on a global all-to-all coupling as a sinusoidal interaction, depending on the phase difference between pairs of oscillators. He better understood the synchronization processes in globally weakly coupled sets of non-linear, almost identical oscillators, finding a phase transition from incoherence to global synchrony and positive feedback between synchronization and effective coupling.
   
Some discussions and experimental results have suggested that many biological systems operate at or close to critical regimes where long-distance correlations support the emergence of collective dynamics despite the local nature of interactions \cite{Mora2011}. However, more is needed to know about the dynamics when interactions are local. Understanding the mechanisms behind emergent behaviour in systems of many interacting units is still an open question \cite{Crutchfield2011}. Even in cases where the dynamics of the isolated units are well known and described, understanding how complex or collective behaviour appears in sets of such units turns out to be a complicated task. There are many examples where collective behaviour arises from the interaction of simple individual units \cite{Strogatz1993}.

A clear example is discussed by Bialek et al. \cite{Bialek2014} regarding travelling flocks of birds, where long-range correlations in the critical regime enable the propagation of information across the flock. In other fields like theoretical biology and neural network studies, arrays of coupled non-linear oscillatory units have proven to be useful models \cite{Mosekilde,Ermentrout1998}. Experimental data regarding brain oscillations also appears to show complex nature \cite{Akay2009}. Collective de-localized phenomena such as the binding problem \cite{Singer2006,Farmer1998}  are verified in a system formed by many interacting units.

In a previous paper, where the model also used in this study was proposed, the conditions under which global synchronization becomes possible when interactions are confined to a local scale were analyzed. Sufficient conditions under which synchronization occurs were established, and insight into how such synchronization happens was reported. Winfree's intuitive idea about the role of interaction with collective rhythms in enabling global synchronization was confirmed, even for systems whose units interact only locally. Positive feedback was found between coupling strength and synchronization at a local scale, the same scale in which the units interact. Simulations suggest that competition among different synchronization centres or communities with different degrees of local synchrony is the key to understanding global synchronization. We also found what we called satellite communities, which indicates that not all communities evolve, at a local scale, in the same way towards global synchronization.

Despite the simple nature of the proposed model, its long time dynamics were found to span over a wide range of behaviours. Even in the simplest case possible, consisting of two coupled oscillators, the system was found to sustain up to six different stationary points in phase space for some specific regions in parameter space. Furthermore, in the limit of large populations, coherent stationary configurations were found to be stable only at a particular region in parameter space, different from the existence region defined by the bifurcation line. The fact that there is a region in parameter space where the coherent solution is available but unstable opens the possibility for more sophisticated collective behaviour.

This contribution focuses on the spatiotemporal behaviours measured through entropic quantities. Such studies have been done previously by the authors on other non-linear systems and have proven their usefulness to characterize the emergence of complexity and chaotic dynamics quantitatively and to identify additional features such as enhanced computation at the edge of chaos \cite{Estevez2015,Estevez2018,Estevez2023}. 

The paper is structured as follows. In Section \ref{Model}, we formally introduced the model and summarized the results of the previous contribution to make this article self-contained. Entropic complexity measures are presented in Section \ref{Measures}, as well as the methods for their estimation. Section \ref{Discussion} shows the results of a detailed complexity analysis of the model in the parameter space. Finally, conclusions are given.

\section{The model}\label{Model}
Consider the model introduced in \cite{garcia22}. It consists of $N$ oscillators arranged in a circle with a Kuramoto-type interaction depending on their phase difference but only with nearest neighbours. The model differs from the classical Kuramoto model \cite{acebron05} not only in the local nature of the coupling but also in a self-term given by an Adler-like oscillator\cite{Adler1973}. The resulting equation of motion for the phases takes the form 
\begin{equation}\label{eq:model}
\dot{\theta}_i=\omega+\gamma \cos(\theta_i)+(-1)^ik\left\lbrace \sin(\theta_{i-1}-\theta_i)+\sin(\theta_{i+1}-\theta_i)\right\rbrace,
\end{equation}
where $i$ $\in [1,N]$ and $N$ is taken to be an even number. $\omega(>0)$ stands for the natural frequency of oscillation for the isolated units, and $\gamma(>0)$ can be seen as a self-feedback or excitability coefficient. The third term corresponds to the local interaction; consequently, $k$ is the coupling strength. Notice that since the sign on this term alternates along the ring, interactions are globally balanced since $N$ is taken to be even, and periodic boundary conditions are imposed.

Adler oscillators, as described by (\ref{eq:model}) with $k=0$, 
\begin{equation}
 \dot{\theta}_i=\omega+\gamma \cos(\theta_i).\label{eq:adler}
\end{equation}
where critical values are found for $\theta^*=\pm \arccos \left ( -\omega/\gamma\right )$, which can only happen if $\gamma\geq\omega$. For $\omega=\gamma$, there is one critical point ($\theta^*=\pi$) which is unstable. It will attract all oscillations starting with a lower $\theta$ phase value and repulsive for all oscillations starting with a larger $\theta$ than the critical $\theta^*$. No critical value is found for $\gamma < \omega$, but the oscillator exhibits intervals for bottleneck, where the speed decreases when the second term is negative. For $\gamma \geq \omega$, two critical points emerge due to a saddle-node bifurcation; one of the critical points will be stable ($\theta^*_1$) and the other unstable ($\theta^*_2$). All trajectories will end ($t\rightarrow\infty$) in the stable critical value.

Equation (\ref{eq:adler}) has analytical solution, for $\omega > \gamma$,
\[
 \begin{array}{ll}
  u_1=\sqrt{\frac{\omega+\gamma}{\omega-\gamma}}\tan \left (  \frac{t}{2}\sqrt{\omega^2-\gamma^2}\right) \\\\
  u_2=-\sqrt{\frac{\omega+\gamma}{\omega-\gamma}}\cot \left (  \frac{t}{2}\sqrt{\omega^2-\gamma^2}\right)
  \end{array}
\]
where $u=\tan(\theta/2)$, and the period of the solutions is $T=2\pi/\omega_A$, $\omega_A=1/2\sqrt{\omega^2-\gamma^2}$, is the Adler frequency.

In \cite{garcia22}, a thorough discussion of the model was reported. No analytical solution can be found when coupling ($k\neq0$) is present. A summary follows for completeness.

For the simplest possible case $N=2$, equilibrium configurations ($\dot{\theta}_1=\dot{\theta}_2=0$) would imply $\cos(\theta_1)=\cos(\theta_2)=a$ given the odd nature of the $\sin(x)$ function. Equilibrium points will therefore appear in two forms, namely $\theta_1^*=\theta_2^*$ and $\theta_1^*=-\theta_2^*$. In the first case, the steady phase value $\theta_1^*$ is given by
\begin{equation}\label{eq:steady1}
\theta_1^*=\pm \arccos\left[-\frac{\omega}{\gamma}\right],
\end{equation}	
with existence condition $\omega\leq\gamma$, rendering two points in the phase space whenever the equality is not true and one whenever it is. In the other case, the steady phase value is given by the solution to the transcendent equation
\begin{equation}\label{eq:steady2}
\cos(\theta_1^*)=\frac{2k}{\gamma}\sin(-2\theta_1^*)-\frac{\omega}{\gamma},
\end{equation} 
which can be rewritten as the quadratic form
\begin{equation}\label{eq:cuadratic}
x \left (\gamma\pm4\sqrt{1-x^2}\right)+\omega=0,
\end{equation}
where $x=\cos \theta_1^*$. Expression (\ref{eq:cuadratic}) results in up to four different solutions. Therefore, up to six different equilibrium points were observed simultaneously, only one being stable but not necessarily globally attractive, namely the positive branch in expression (\ref{eq:steady1}).

For the case $N \geq 4$, systematic determination of all possible steady configurations poses a challenging combinatorial problem. Nevertheless, a similar analysis can be performed, noticing that the configuration corresponding to $\theta_i=\arccos[-\omega/\gamma]$,  $\forall i \in [1, N]$ is still a possible steady state. This configuration was numerically found to be stable for all even $N \in [4,500]$ in a specific region of parameter space, which opens the possibility for richer dynamical behaviours.

Finally, equation (\ref{eq:model}) can be compared to the closely related model discussed in \cite{Estevez2018,Estevez2023}, where the same type of Adler equation is used, but the coupling is taken by the absolute value of the phase of the neighbouring units, the so-called Alonso model. In this other model, regions of complexity were identified together with regions of chaotic regime.

\section{Complexity measures}\label{Measures}

Complexity can be characterized in several different ways. However, the emergence of what can be considered complex behaviour is as far as possible from two extreme cases: completely random configuration and entirely predictable pattern-driven configuration (e.g. periodic). The fact that both extremes must be avoided struck a compromise of how much complexity the system's dynamics can accommodate given an initial configuration. 

In the above sense, complex configurations are a non-trivial mixture of random and pattern-forming behaviour. Non-trivial means that mesoscopic spatial correlations should be a feature of the system. In this context, mesoscopic should be understood as larger than local (just a few units) and less than global correlation. In the time dimension, a similar analysis is still valid; we add the fact that in a chaotic regime, there is a high sensibility to change in the initial conditions, while in a time predictable system, we will assume that two initial configurations differing in just a few units, will evolve in such a way that their difference does not grow excessively fast in time (e.g. exponentially fast). 

To measure randomness, entropy density $h$, defined through Shannon entropy, can be used \cite{Cover2006}. Consider a bi-infinite sequence of $\Lambda=\ldots \lambda_{-2},\lambda_{-1},\lambda_{0},\lambda_{1},\lambda_{2},\ldots$, and a subsequence of length L as $\lambda^L$. Let us define the Shannon block entropy as 
\begin{equation}
H_{\Lambda}(L)=-\sum_{\lambda^l \in \Lambda^L}p(\lambda^L)\log p(\lambda^L),
\end{equation}
where the sum runs over all possible subsequences $\lambda^L$ of length $L$ taken from the set of subsequence $\Lambda^L(\subset\Lambda)$ of length $L$. $p(\lambda^L)$ is the probability of a particular sequence $\lambda^L$. The entropy density is defined as 
\begin{equation}\label{eq:entropy}
h= \lim_{L \rightarrow \infty}\frac{H_{\Lambda}(L)}{L}.
\end{equation}
It measures the production of information in a sequence as it is parsed in a given sequential order. It can also be interpreted as the irreducible randomness of a bi-infinite sequence once all sources of predictability have been considered.

To measure correlation and consequently pattern formation, effective complexity measure ($ECM$), also known as excess entropy $E$, will be used \cite{Grassberger1986,Crutchfield2011}, defined as the convergence of the defect of the entropy density \cite{Crutchfield2013},
\begin{equation}\label{eq:excess}
E=\sum_{L=1}^{\infty}[h_L-h],
\end{equation}

where $h_L=H_{\Lambda}(L)-H_{\Lambda}(L-1)$. ECM of an infinite system is the mutual information between the two halves of the system; $ E$ can also be seen as the intrinsic redundancy or apparent memory of a symbol source \cite{Crutchfield2013}. EMC measures the convergence rate of a finite-size estimate of entropy density to its actual value. In other words, it measures, for each $L$ value, how much we are overestimating the randomness of the information source given that we are only taking into account $L$ different observations up to that point. Thus, its total value directly measures correlations at all possible scales. 

In that sense, complexity will only be compatible with intermediate values of both entropic measures, leaving space for non-trivial behaviours at larger scales than the natural scale of interactions. 

Finally, sensitivity to initial conditions will be measured by the information distance $d$, which follows from Kolmogorov randomness. Information distance between two systems measures the length of the smaller algorithm that can transform one system description into the other \cite{Li2004a}.
\begin{equation}\label{eq:kdistance}
d_{\Lambda,\Lambda'}=\frac{K_{\Lambda \Lambda'}-\min \{K_{\Lambda},K_{\Lambda'}\} }{\max\{ K_{\Lambda},K_{\Lambda'} \}}.
\end{equation} 
$K(s)$ is the length of the smaller algorithm that reproduces the string $s$ in a Universal Turing Machine. The distance defined by (\ref{eq:kdistance}) is not a Hamming-type distance because it does not measure the number of different bits between two sequences. Instead, it indicates how innovative one sequence is once the other sequence has been considered.

\subsection{Simulation and methods}

We are interested in a discrete spatial variant of the original oscillator problem. As discussed in the seminal paper of Winfree \cite{Winfree1967}, since the phase value $\theta$ is not one of the observable, a periodic function of the phase, experimentally observable, must be introduced. Instead of using a triangular wave as used by Winfree, Alonso \cite{Alonso2017} introduced $\sin \theta$, which he called activity, as the observable function. The activity has been further used in other studies of a similar system of coupled Adler oscillators \cite{Estevez2018,Estevez2019,garcia22,Estevez2023} and will also be used here. The choice makes the observed magnitude between $0$ and $1$, which is convenient for the discretization procedure explained below without loss of information. Two oscillator phases on the circle can come as far away as $\pi$, which means, for the activity, a change of sign for the same absolute value. 

The name of the activity comes from the fact that in the uncoupled Adler equation, $\dot{\theta}$ is proportional to $\cos \theta$, and following the latter amounts to following the phase speed. In a coupled system, such as the Kuramoto model \cite{kuramoto75}, the complex order parameter is introduced as
\begin{equation}
 r e^{i \psi}=\frac{1}{N}\sum\limits_{j=1}^N e^{i\theta_j},\label{eq:order}
\end{equation}
$\psi$ is a mean phase towards which the phase of the individual oscillators $\theta_i$ is pulled. On the other hand, $r$, the effective strength of the coupling, is related to the mean value of $\langle e^{i\theta}\rangle $ as shown by Strogatz \cite{Strogatz00}, which can be written in terms of the activity. Following the behaviour of the activity, as in the uncoupled system, makes physical sense. 

A similar analysis was performed in the model we are discussing \cite{garcia22}, where local $r_k$ and $\psi_k$ parameters are introduced using a similar expression as (\ref{eq:order}), but summing over the coupled units. It has been shown that the (lack of) collective behaviour can be followed by the evolution of $r_i$ over time and, from there, the achievement or not of synchronization. The reader can refer to this previous study for details.

The activity value is discretized in the following way: For each $\theta$ vector, the mean value of the activity is calculated; then, for each unit, if the activity value is above the mean, it is labelled by a $1$, otherwise by a $0$.
All numerical computations were performed with an error of up to $10^{-6}$.  

At any given time step, the density $\rho$ will measure the fraction of symbol $1$ in the vector. Complexity variations followed by the entropic magnitudes can be associated with two processes: erasing one symbol at the expense of the other and shuffling existing symbols. At the end of the data reduction process, the system's state at a given time is characterized by a binary vector of length $N$. 

Much literature describes different estimation methods of the entropic magnitudes from finite sequences \cite{Grassbergerb,Rapp2001,Lesne2009}. Here an approach based on Lempel-Ziv complexity will be used as described in Estevez et al. \cite{Estevez2015}. Expressions (\ref{eq:entropy}), (\ref{eq:excess}) and (\ref{eq:kdistance}) are defined for infinite sets of data, and estimations of both quantities are needed from finite size data. All entropic measures were calculated using the same software as in ref. \cite{Estevez2015}.

Complexity analysis will be done of the spatiotemporal behaviour of the system, considering the ring of oscillators as a $N$ dimensional vector starting at an arbitrary unit which also fixes the ending unit. A $T\times N$ array, named the spatiotemporal map, is constructed by taking fix time step and considering the evolution of the $N$-vector up to $T$ time steps. 

The differential equations were solved using a Runge-Kutta (4,5) method as implemented in the GNU Scientific Library (GSL). At random instances, the results were compared with those obtained using Mathematica and no significant discrepancies were found. All simulations were performed with $N=5 \times 10^{2}$; the system was left to evolve, starting at a uniformly distributed random configuration, for $T=2000$ time steps with a fixed difference $\Delta t=1$ between them. The numerical method was used with an error of $10^{-6}$.

All spatial measures, namely $h$ and $E$, were determined for the last $100$ configurations in $10$ different runs for a total set of $1000$ values, and average values are reported for each combination of parameters. Temporal entropic measures such as $h_t$ and $E_t$, on the other hand, were determined from the temporal evolution of each unit, and the average value over the entire system was taken. Simulations were performed for the information distance calculations, modifying a single unit in the initial conditions. The distance was computed using the average values over the last 100 time steps in 10 different runs; the value of $d$ between reference and perturbed simulations where determined at each iteration. 

In what follows, $k$ is fixed at $1$, and the parameter space is determined by $(\omega,\gamma)$.

\section{Results and discussion}\label{Discussion}

\begin{figure}[ht!]
	\centering
		\includegraphics[scale=0.6]{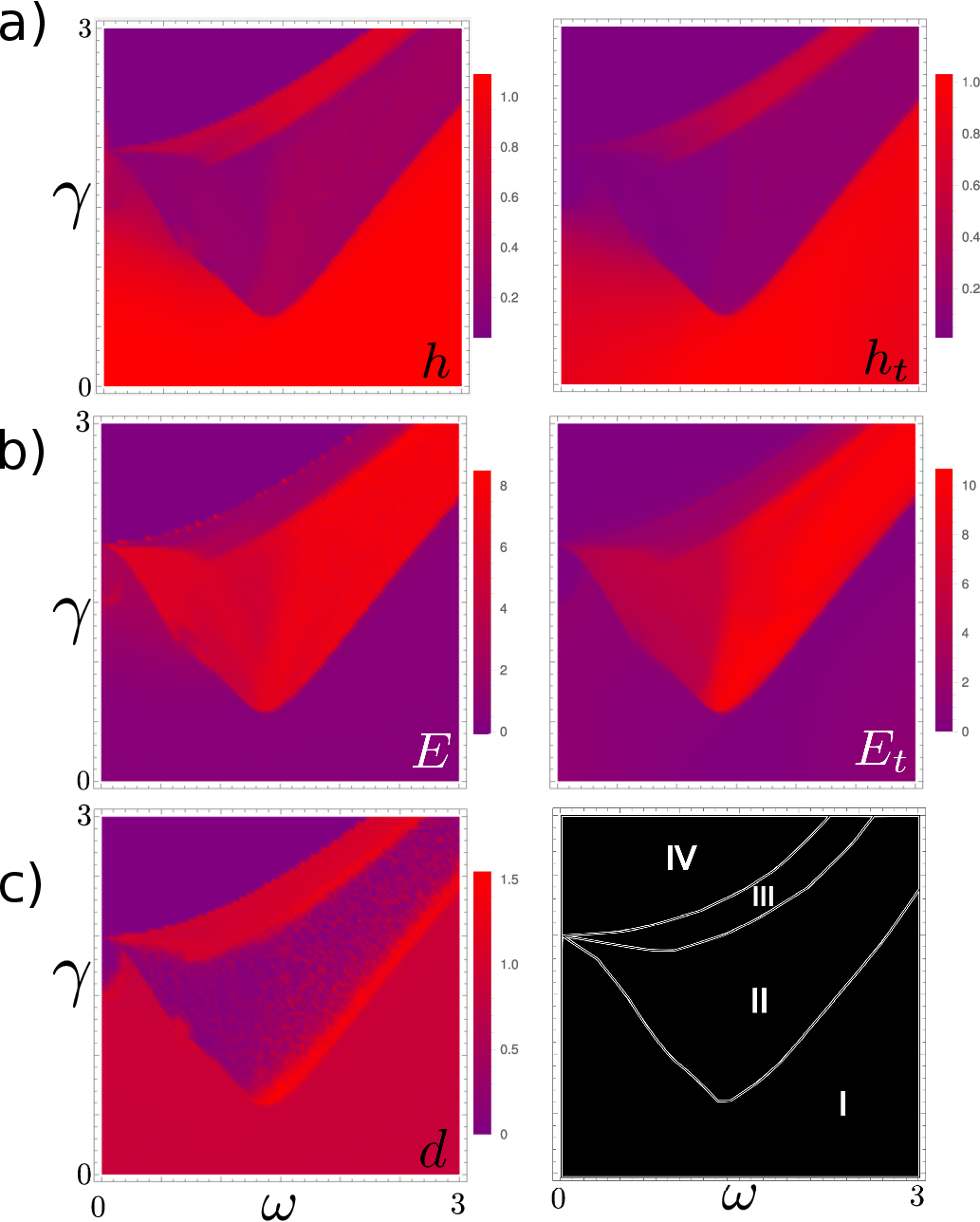}	
		\caption{Maps of the entropic magnitudes in the parameter space $(\omega,\gamma)$. (a) Entropy density can be taken as a measure of the information production of the system and, therefore, unpredictability or irreducible randomness. (b) Effective complexity measure (ECM) or excess entropy $E$ ($E_t$) measures the correlation at all length scales in the system. ECM is the mutual information between two halves of a bi-infinite string of symbols. In our case, the string of symbols is the binarized value of the activity $\sin(\theta)$. Larger values of ECM signal better structuring of the data and emergence of patterns. (c)left is the information distance map between two systems with the same (random) initial condition except for one oscillating unit at the centre. For the right column, the sequence of values is taken along the oscillating units. The magnitudes are computed for 500 units, evolving 2000 time steps, and the values of $h$, $E$, and $d$ are calculated. This is repeated for ten runs, and mean values are computed. For the left column, the sequences are taken along the time evolution for each oscillating unit, and mean values are reported. (c) right is a diagram of the identified regions from the entropic plots. Region IV will be called absorbing, where the $h$ and $E$ are near zero; region III shows intermediate values of $E$ and $h$, which is taken as an indication of complex behaviour; the same can be said of region II, where $h$ is even lower, but not zero, while $E$ increases respect to the previous two regions. Finally, region I is chaotic with a maximum value of $h$ and $d$, while $E$ is around zero.\label{fig:Figure_I}}
\end{figure}
 
 \textbf{Regions in the parameter space}. We identified four regions following Figure \ref{fig:Figure_I}. The region labelled as I (right \ref{fig:Figure_I}c) has a high entropy density, low $E$ and high sensitivity to initial conditions, which bears all the features of a chaotic regime. Region IV exhibit near zero entropy, excess entropy, and near zero $d$ value. This region has been labelled absorbing by \cite{Alonso2017}. A similar region was observed in the model discussed in \cite{Estevez2015}. In this region, as in the cited model, density value $\rho$ shows that this behaviour results from a heavy erasure of one symbol at the expense of the other, resulting in an evolution to constant spatial vectors. Global phase synchronization is, therefore, achieved. The region of parameter space where this is verified roughly matches the stability region for the stationary globally coherent solution previously observed for this model \cite{garcia22}.

Region II is triangular, with a cusp at $(\omega, \gamma)\approx (1.4,0.6)$. The region exhibits intermediate entropy density values with large excess entropy in both spatial and temporal magnitudes. Information distance shows a granular image with intermediate values compared to regions I and IV. Another interesting feature is the bright $d$ band of this region's lower right-side boundary. Similar characterist are found in Region III, although the $h$ values are slightly higher than in Region II. According to the entropic values, regions II and III are where complex behaviour should be expected.

Regarding noise sensitivity, the information distance map in \ref{fig:Figure_I}c-left shows that regions I ad III are rather sensitive to the initial conditions. As a potential computing device, operation in this region faces the formidable challenge of lack of robustness for any system disturbance. Yet, in all cases, the nature of the system dynamic does not change with noise. The mean values of the entropic measures, entropy density, and ECM computed from $1000$ values showed small variations as exposed by the standard deviation value not larger than $0.02$ for the entropy density and $0.05$ for the EMC in regions I and III.

\begin{figure}[ht!]%
	\centering
		\includegraphics[scale=0.5]{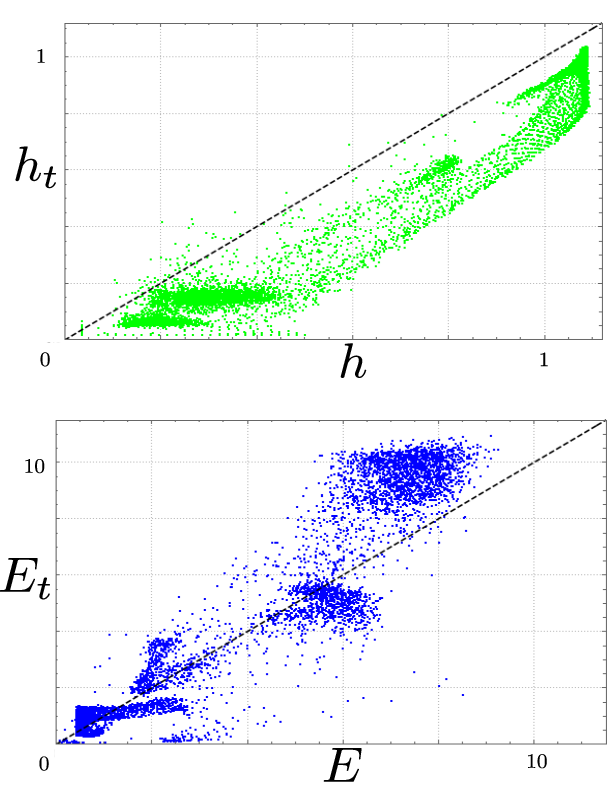}	
		\caption{Plots of the relation between the spatial and the time behaviour of the entropic magnitudes $h$ and $E$. They have no functional dependence, yet clear trends can be followed. \textbf{(upper)} The entropy density $h_t$ vs $h$. Two features can be seen; almost all points are below the diagonal line $h_t=h$, which indicates that a given value of spatial entropy density $h$ can accommodate less temporal entropy density $h_t$. The second feature is a trend where increasing values of $h$ correspond to increasing values of $h_t$. \textbf{(below)} The effective complexity measure also does not show a functional dependence between the temporal behaviour $E_t$ and the spatial behaviour $E$. For a given value of $E$, the system can accommodate a range of time-effective complexity measure $E_t$ values. This range spans below and above the main diagonal. As in the case of the entropy density, there is a trend that larger values of $E$ result in larger values of $E_t$. \label{fig:Figure_II}}
\end{figure}

Comparison between spatial and temporal magnitudes allows exploring the correlation between both behaviours as shown in Figure \ref{fig:Figure_II}. No functional relation exists between the entropic measures at the spatial and temporal dimensions. It must be noticed that for a given value of $h$, there is a range of $h_t$ values. Therefore, different temporal disorder levels are allowed for a given spatial disorder. 

For the $h_t$ vs $h$ plot, almost all points lie below the diagonal line ($h_t=h$), which means that spatial disorder can accommodate, at most, a lower range of values for the temporal disorder. The plot of excess entropy shows that the diagonal line is not of particular significance. Spatial pattern formation, as described by $E$, can accommodate a large range of temporal patterns measured by $E_t$, yet in general, the larger $E$, the higher the range of $E_t$ values.  

Two considerations can be made from both plots in figure \ref{fig:Figure_II}. First, there is a correlation between spatial and temporal behaviour, which should come as no surprise, as equation (\ref{eq:model}) establishes a link between the oscillating units, but, in any case, this is by no means trivial because the interaction between units is local, and the plotted magnitudes are global. The fact that there is a clear tendency of increasing $h_t$ values for increasing $h$ values points to this loose correlation.

The second consideration comes from the lack of functional relation in both plots of Figure \ref{fig:Figure_II}, pointing to the fact that the whole system dynamic is not captured by either the entropy density or the excess entropy when taken alone. There is a relation between how much a given disorder allows the structuring of the system, both in the spatial and temporal dimensions, but this is not the whole story. The fact that for given values of $h$ or $E$, there is a range of values of $h_t$ and $E_t$, respectively, shows that the same amount of disorder can accommodate different levels of structuring. Intermediate values of $h$, $h_t$ and $E$, $E_t$ are interesting in terms of complex correlations that could give rise to complex spatiotemporal behaviours.

\begin{figure}[ht!]%
	\centering
		\includegraphics[scale=0.5]{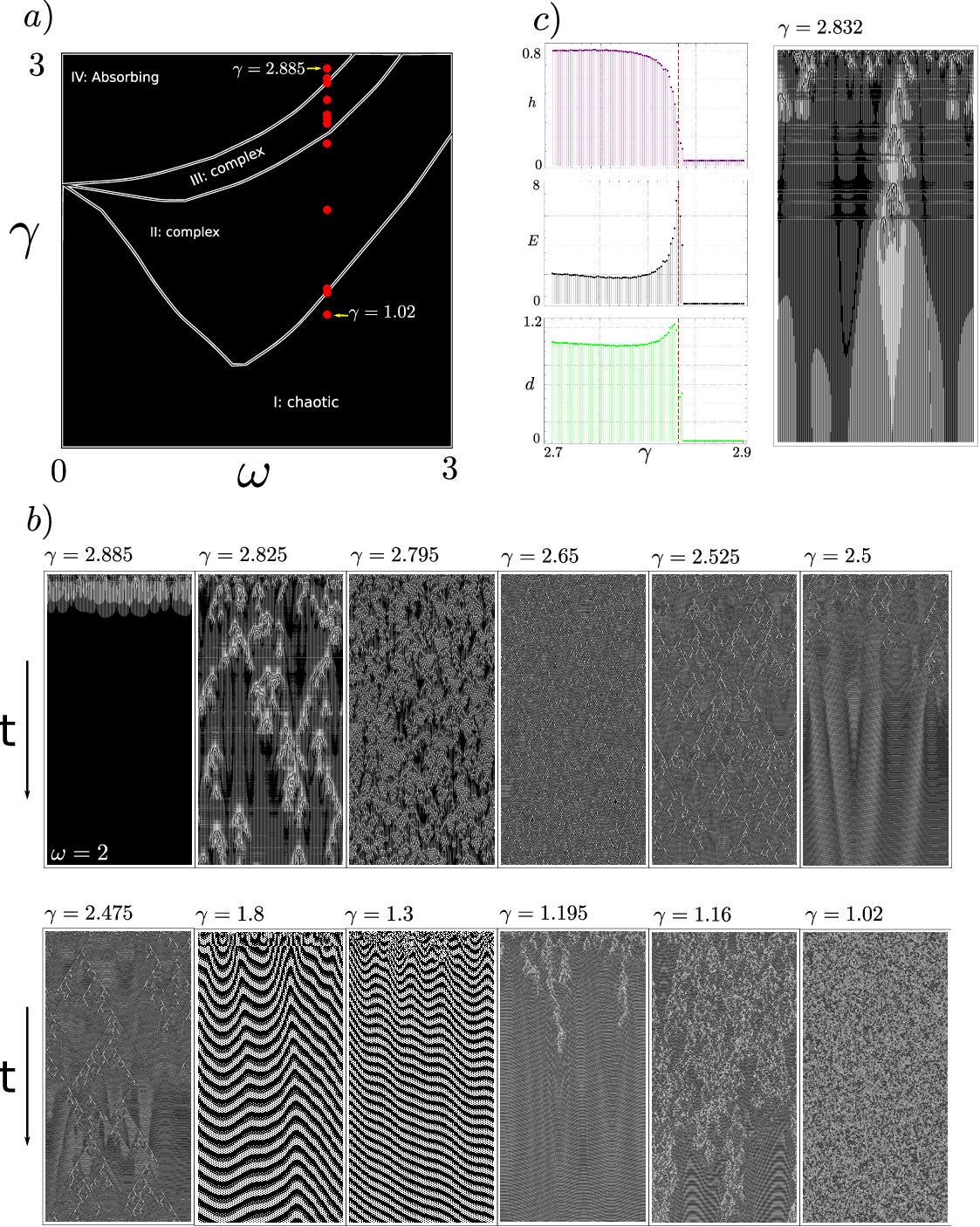}	
		\caption{\textbf{a)} Schematic representation of the parameter space where the points of the spatiotemporal map analysis are shown as red dots. \textbf{b)} Spatio-temporal maps corresponding to the red points depicted in (a). Time is the vertical axis, while the horizontal axis runs through the oscillators. In every case, a random initial configuration was taken. The total number of time steps is $1000$. $\gamma=2.885$ is in the absorbing region; $\gamma=2.825,2.795$ are at the boundary between the absorbing and complex region III; $\gamma=2.650,2.525,2.500,2.475$ are in the complex region III; $\gamma=1.800,1.300$ are in complex region II; $\gamma=1.195, 1.160$ are at the boundary between complex region II and the chaotic region; $\gamma=1.02$ is well within the chaotic region. \textbf{c)} behaviour at the boundary between the absorbing and complex region III. \textbf{left}: The behaviour of the entropic magnitudes as the boundary is crossed. Observe the spike in $E$, the excess entropy, at the same $\gamma$ value where $h$ and $d$ drop. In CA and Alonso models, this has been taken before as a sign of enhanced computation at the edge of chaos \cite{Estevez2019}. \textbf{right:} The spatio-temporal map at $\gamma=2.832$ in the boundary between the two regions. Long, time and spatially correlated behaviour can be observed.\label{fig:Figure_III}}
\end{figure}

\textbf{Spatio-temporal maps}. The above discussion can be further enriched by observing the spatiotemporal evolution of the system for specific $(\omega,\gamma)$ values. As already described, a matrix can be constructed from the binarized spatial vectors as they evolve at fixed time steps. In the spatiotemporal diagrams, rows represent the spatial dimension, while columns represent the time evolution from the top (initial condition) to the bottom (final spatial configuration). Matrices like this are shown in Figure \ref{fig:Figure_III} b). The black (white) pixels correspond to $1$ ($0$) states.

The matrices shown correspond to different points on parameter space, all with $\omega=2$, as depicted in the diagram (a) of Figure \ref{fig:Figure_III}. Each circle correspond to a spatiotemporal matrix shown in (c); the value of $\gamma$ eliminates any ambiguity between circles in figure \ref{fig:Figure_III}a and the corresponding spatiotemporal matrix of figure \ref{fig:Figure_III}c. 

In the case of $\gamma=2.885$, a transient period is present, after which the system falls into a stationary coherent state, as expected for region IV. As one moves closer to the complex region III, the transient period becomes larger, but the final states remain defined by $\theta_i=\arccos(-\omega/\gamma)$ $\forall i \in [1, N]$, independently of the starting configuration. The coherent state remains a global attractor. All possible starting conditions fall in its basin of attraction, and a coherent final state is achieved. The transient gets larger as we get closer to the boundary between regions IV and III, indicating that the system wanders in phase space before settling into its attractor. 

As the boundary is crossed, the coherent state is no longer globally attractive, and the system stops evolving towards coherent steady states. The space-time matrix corresponding to $\gamma=2.825$, located at the frontier between the absorbing region IV and the complex region III, features the appearance of unseen structures. These new patterns propagate in space and are sustained over a time interval. These patterns travel in a background exhibiting a wavy picture. It is unclear if the initial conditions persist, at least partially, as time evolves. However, some travelling patterns initiate from the initial configuration, and some appear along the way. The spatiotemporal map closely resembles those seen in CA with improved computational capabilities \cite{wolfram86,wolfram02}. Slightly below, at $\gamma=2.795$, the density of these structures is noticeably bigger, to the point where they almost occupy the entire matrix. Uninterrupted paths through these structures can be traced from units in the initial configuration through the time scale.

As we move into complex Region III, the nature of these structures changes. Notice how, for $\gamma=2.650$, more compact repetitive structures appear. Triangular patterns are common, indicating highly correlated states in both spatial and temporal dimensions. The scale of these correlations is even bigger for $ \gamma=2.525 $, with a space-time matrix featuring longer and clearer lines and even bigger triangular patterns. These patterns highly resemble those observed on complex (W3) CA. It was found that the particular spatiotemporal map depends on the initial configuration, while the general features remain the same. Already for this $(\omega,\gamma)=(2, 2.525)$ point, an uninterrupted path through these emerging features can be traced from the initial condition all the way down in the whole time range.

An even more interesting behaviour emerges at the very core of the complex region III. For $\gamma=2.500$, a clear wavy pattern appears. The previously observed structures are still present. In some cases, as shown in the corresponding spatial-temporal map, the initial patterns die after some time; in other instances, they keep evolving through the whole studied time range. Again, this shows that such structures are dependent on the initial conditions. As discussed below, the wavy pattern is witnessing the appearance of local collective behaviour. 

At $\gamma=2.475$, the spatiotemporal map shows both the wavy and the triangular patterns sustained over time. Again, an uninterrupted path can be followed in the whole temporal range through the triangular features. As the system moves from complex region III to complex region II, only the wavy patterns prevail, as seen for $\gamma=1.800$ and $1.300$. $\gamma=1.800$ closely resembles a similar map in the needle region of the Alonso model as reported in \cite{Estevez2023}. At $\gamma=1.300$, the ''wavelength'' has decreased, and the ''frequency'' has increased compared to the previous point, again resembling a similar behaviour in the needle region of Alonso's model. 

Finally, as we get closer to the chaotic region from Region II, the initial randomness starts to creep into the map for longer times, as seen on the matrix corresponding to $\gamma=1.195$ located at the exact border between these regions. In this frontier, the wavy patterns are still observable, as further verified at $\gamma=1.16$. However, randomness eventually takes over the entire landscape, as observed for $\gamma=1.02$, already in the chaotic regime.

Some authors have discussed the possible relations between complexity and bifurcation theory \cite{Alonso2017}. Our model sustains at least one bifurcation when crossing the diagonal $\gamma=\omega$, from which the globally coherent state becomes available but not necessarily stable \cite{garcia22}. Diagonal line $\gamma=\omega$ falls right inside our complex regions. It is interesting, however, that in our model, complex behaviour is observed on both sides of the diagonal. Complex dynamics appear with and without coherent steady states on the $n-$dimensional phase space. Since existence conditions for all possible steady configurations have not been found, there is no certainty about any other kind of steady state when complex behaviour is observed.

\textbf{EOC}. The system was explored to find regions of possible compliance with the edge of chaos hypothesis (EOC). EOC poses the possibility of a relationship between enhancement of the computational capabilities of a given system as it transitions to chaos \cite{Langton1990}. Since proposed, this idea has been widely discussed and verified in various non-linear systems \cite{packard88,su89,kauffman92,beggs08}. Complex regimes in our parameter space could be seen as a natural consequence of the system's transition from absorbing to chaotic regimes. 

A possible verification of the EOC hypothesis was found. Figure \ref{fig:Figure_III} c) shows a sudden increase in excess entropy, as $\gamma$ is changed, right at the verge of entering the complex region III, which reportedly has a high sensitivity to initial conditions as evidenced by the high values in $d$. This means that, potentially, the system has a better computational performance at $\gamma=2.832$, at which the sudden peak appears, compared to the rest of the $\gamma$ values in the complex region III. A similar result in the Alonso model and CA has been previously reported \cite{Estevez2019}. This is not evident when looking at the spatiotemporal matrix corresponding to $\gamma=2.832$ in the same figure. Yet, $E$ captures a peak in correlations or system spatial memory at this critical value while there is a sudden drop of $h$. Interestingly, this enhancement in computational capabilities is not taking place on the edge of chaos in the sense of randomness but rather on the edge of a complex, highly-sensitive regime as region III.

For Region II, there is also a certain amount of pattern formation and memory, which, combined with fluctuating sensitivity levels, allows the system to handle information to a certain level. We can approach complexity discussions in terms of computational capabilities, which, in turn, can be analyzed by looking at the system's capacity to perform some basic tasks on the information it receives from the environment. Note that, from an anthropocentric point of view, these basic tasks could be taken as the storage, transmission, and modification in some meaningful way of the received information. To be capable of doing so, the system can not be in a dynamically trivial or chaotic regime. In the former, any initial information is lost since it is completely modified as the system evolves towards its attractor. The same final state is reached no matter what the starting condition is. All history is lost. Initial information is also lost in the latter due to the destruction of correlations at all scales. Small changes in the information entering the system are amplified, causing different responses. No meaningful computation can be performed. 


\textbf{Rotation number}. In an attempt to relate our complexity analysis, based on entropic measures, with some quantities familiar to coupled oscillator theory, we can turn to the rotation number. The rotation number is defined as \cite{Ariaratnam2001}

\begin{equation}
\varsigma_i=\lim_{t \rightarrow \infty} \varsigma_i(t)= \lim_{t \rightarrow \infty} \frac{\theta_i}{t}
\end{equation}

\begin{figure}[ht!]
\centering
		\includegraphics[scale=0.8]{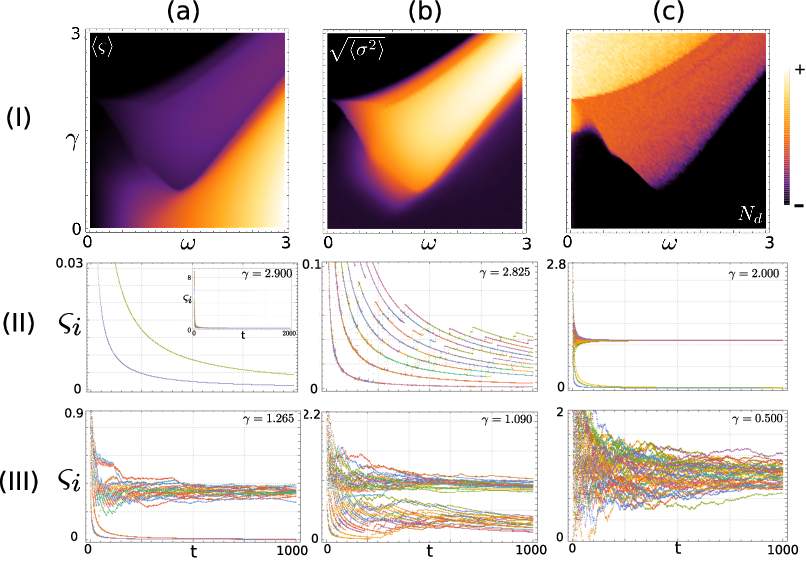}	
		\caption{The rotation number $\varsigma$ can be considered an effective frequency for the oscillators used to study synchronization processes. \textbf{(I) a)} The mean value of the rotation number $\langle \varsigma \rangle$ was calculated from the set of oscillators. The map, in parameter space, roughly describes the same regions as those identified in Fig. \ref{fig:Figure_I}. Observe that in the absorbing region, the average value is zero; this also follows the behaviour of the individual oscillating units. In such cases, the oscillators are said to be ''dead''. The chaotic region I show the larger values of $\langle \varsigma \rangle$, while the complex regions exhibit intermediate values. \textbf{(I) b)} The average standard deviation $\sqrt{\langle \varsigma \rangle}$ map of the rotation numbers, in parameter space, shows that the largest variation can be found in the complex regions. In contrast, the absorbing and chaotic regions have less spread values with respect to the mean $\langle \varsigma\rangle$. \textbf{(I) c)} The map of the number of dead oscillators in parameter space, with an error of 0.002. In the absorbing region, all oscillators eventually die. There are no dead oscillators in the chaotic region, while an intermediate value was found in the complex region. \textbf{(II, III)} Temporal evolution of the rotation number $\varsigma_i$ at different points of the parameter space, for $50$ different, randomly chosen, individual oscillating units. $\omega$ was kept fixed at a value of $2$ while $6$ different values of $\gamma$ were chosen, corresponding to different regions. All oscillators eventually die in the absorbing region ($\gamma=2.900$) (see the inset), although two distinct curves can be identified. $\gamma=2.825$ is at the boundary between the absorbing and the complex III regions, where enhanced computation was reported in Fig \ref{fig:Figure_III}. $\gamma=2.000$ is at the boundary between both complex regions; $\gamma=1.265$ is within the complex region II; $\gamma=1.090$ is at the boundary between complex region II and the chaotic region while; finally, $\gamma=0.500$ is well within the chaotic region.\label{fig:Figure_IV}}
\end{figure}

This number can be interpreted as an effective or average frequency and is especially useful for studying frequency synchronization: two frequency-locked oscillators will have equal rotation numbers. It must be noticed, however, that the rotation number fails to capture some long-term dynamics, such as the oscillation of phases around some mean value. Furthermore, $\varsigma_i(t)$ cannot distinguish between an intermittently accelerated evolution of individual phases and an oscillatory one. This is one of the main advantages of the local order parameter approach used in \cite{garcia22}, where such oscillations were observed. In our simulations, the 500 oscillators system was integrated throughout 2000 time steps, and the final $\varsigma_i(t)$ value was taken as a good limit estimator. Then, to analyze the behaviour of the rotation number as a function of the model parameters, the average was taken over all units in the system. A rotation number map was constructed with the average value  $ \langle\varsigma_i\rangle$. The results are shown in Figure \ref{fig:Figure_IV} (I) a).

The largest average rotation number values are reported for the chaotic regime, indicating that, in this region of parameter space, the oscillators are, on average, faster than in any other region. Secondly, in both complex regions (II and III), relatively low values of rotation number are calculated. In these regions, the oscillators' evolution occurs at a slower rate. Finally, practically zero values of rotation number are seen for the absorbing region indicating the death of oscillations for long times, which is consistent with the fact that entropic indicators suggest that the system evolves to stationary spatially trivial states. 

Analyzing the behaviour of the standard deviation around the reported average values is useful. Figure \ref{fig:Figure_IV} (I) b) shows the standard deviation of the mean value of the rotation number. The complex regions are the ones that show higher standard deviation values, indicating a greater dispersion around the mean value. In the case of the chaotic region, the standard deviation values are much lower, indicating a higher similarity between individual values of rotation number. Finally, the absorbent region shows null standard deviation values meaning that all units behave similarly. All units reach the same stationary phase value, thus resulting in the death of oscillations.

To clarify the processes in each case, we analyzed the number of units experiencing the death of oscillations in each region. Every oscillator for which $\vert \varsigma_i \vert \leq \epsilon$ is taken as dead. Thus, Figure \ref{fig:Figure_IV} (I) c) shows, with an error of $\epsilon = 0.002$, the number of dead oscillators in each region after 2000 iterations or time steps. Notice how extreme behaviours are observed in the absorbent and chaotic regions. Most units die for long times in the absorbent region, while no unit dies in the chaotic regime, at least not on this timescale. However, the complex regions exhibit a more heterogeneous behaviour for the frequencies. In these regions, a smaller fraction of the total population dies, leaving a standing question about the relation between the effective frequencies of the rest of the units in the system.

The analysis of Figure \ref{fig:Figure_IV} (II) and (III) answers such a question. The procedure for constructing the plots was to determine the individual values of the rotation number $\varsigma_i (t)$ for all system units along the first 1000 iterations. The process was repeated for different points in the parameter space belonging to the identified regions. The plots show a random selection of 50 units.

Notice how, for parameters ( $\omega$, $\gamma$)=($2.0$, $2.9$) on Figure \ref{fig:Figure_IV} (II) a) belonging to the absorbing region, all individual rotation numbers decay at roughly the same rate until it is practically null for all displayed units (the inset shows the effective dead of all oscillators after sufficient time). This is consistent with all previous reasoning about the dynamics of this region. However, for $\gamma = 2.825$, at the boundary between the absorbing region and the complex III, a change in this behaviour is observed, as shown in Figure \ref{fig:Figure_IV} (II) b). Individual units slow down at different rates to the point where some do not die. This could directly relate to the type of structures observed in complex Region III.

As we move to complex region II, units in the system show two different behaviours, as can be seen in Figure \ref{fig:Figure_IV} (II) c) for $\gamma=2.000$. A fraction of the total population dies, but the rest of the system now shares a common non-null value of the rotation number $\varsigma_i(t)$. The system enters a partially phase-locked state, where some units stop oscillating, and the rest oscillates at the same effective frequency. In terms of the phase circle, the above can be understood as a group of oscillators reaching a steady state while the rest travel the circle at a single pace. This would explain the formation of long-term structures in the complex regime since the relations established in the phase space between the different oscillators are maintained over time.
 
In the phase circle, two fundamental groups of oscillators start to be observed, one faster than the other. If we continue to decrease $\gamma$, approaching the chaotic regime, dispersion increases in the phase-locked group as observed at $\gamma=1.265$ in Figure \ref{fig:Figure_IV} (III) a). Eventually, as shown in Figure \ref{fig:Figure_IV} (III) b) for $\gamma=1.090$, entering the chaotic region, we observe that a portion of oscillators that died in the complex region II start reaching small non-null values of individual rotation number and the fraction that used to be phase-locked starts to abandon the locked regime. 

Finally, as shown for $\gamma = 0.500$ in Figure \ref{fig:Figure_IV} (III) c), at the core of chaotic behaviour, all units are grouped around a single effective frequency value but without achieving the frequency-locked state. Consequently, the structures or correlations that may appear between the different oscillators at a particular moment cannot last in time. Likewise, the fact that each unit evolves at a different rate causes the initial condition to heavily influence how the system evolves.

\begin{figure}[ht!]
	\centering
	\includegraphics[scale=.6]{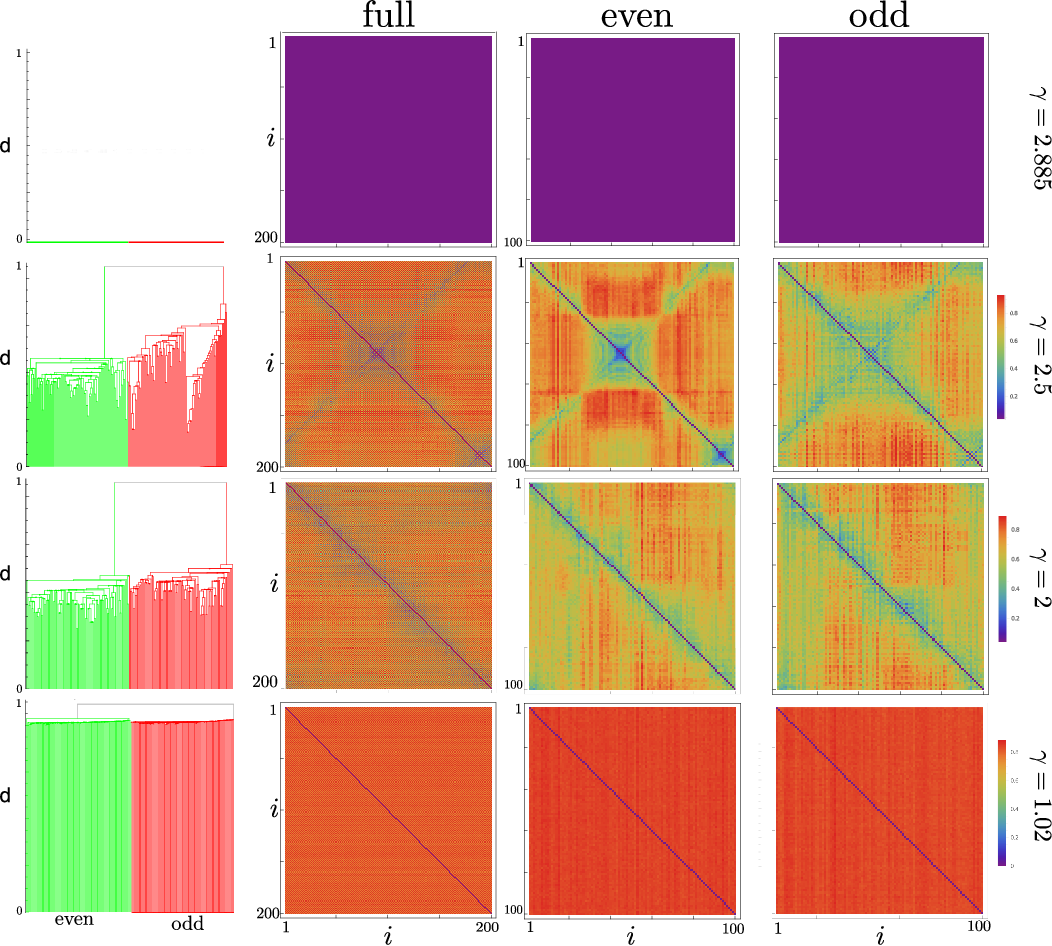}	
	\caption{ Dendrograms (left) are obtained from the algorithmic distance matrices between all units in the system (left) for different points in parameter space ($\omega=2.0$). The dendrograms were built using the smallest intercluster dissimilarity criteria. Distance matrices are shown from columns 2-4, further decomposed in odd and even contributions (third and fourth column). For the absorbing region ($\gamma=2.885$), all units are collectively synchronized, and the distance is zero (first row); for the chaotic region ($\gamma=1.02$, fourth row), there is no correlation between the behaviour of the oscillating units and the distance between them is maximum. $\gamma=2.5$ is in the complex region II, and the hierarchy of clusters can be seen in the dendrogram, with the appearance of local communities and relations at several levels (second row). At the boundary between the complex region II and III ($\gamma=2.00$), clustering is still a feature (third row). However, the levels of relations between the different clusters are more shallow than in the previous (second row) case. \label{fig:Figure_V}}
\end{figure}

\textbf{Distance matrix}. Further evidence of collective behaviour in the complex regions can be found in the algorithmic space. The time evolution of any given unit can be algorithmically compared with any other unit by the computation of $d$ between the corresponding sequences. A distance matrix can then be obtained for any point in parameter space, as shown in Figure \ref{fig:Figure_V}. Independent units would correspond to maximum values in all non-diagonal positions in such a matrix (bottom row in Figure \ref{fig:Figure_V}), while related or redundant units would correspond to a null matrix (top row in Figure \ref{fig:Figure_V}). Any other kind of matrix would indicate correlated evolution between different units. That is the case for $\gamma=2.000$ and $\gamma=2.500$ in Figure \ref{fig:Figure_V}. Intermediate values of $d$ outside the diagonal indicate non-trivial correlations between different population sites. 

Such distance matrix can be used to build dendrograms showing the hierarchies between oscillating units. The dendrogram shows the maximum distance between the oscillating units for the chaotic region, as seen for $\gamma = 1.020$ in Figure \ref{fig:Figure_V}. In the absorbent region ($\gamma=2.885$), the dendrogram shows the minimum distance between the oscillating units. Within complex regions (both I and II, $\gamma=2.000,2.500$), the information distance between the units is in the range of the intermediate values. However, more interestingly, a distance hierarchy appears with the local clustering of the units. This is direct evidence of local correlations in the time evolution of individual units. Off diagonal patterns in the distance matrix is evidence of such local correlations. The observation of these dendrograms seems to confirm the more complex nature of the correlation between the oscillating units. Several levels in the dendrogram mean there is a hierarchy in local communities, and some long-range correlation is built upon lower-scale correlations. Overall, there are local communities that remain correlated on time. 

The introduction of an order parameter in the original Kuramoto model, with global interaction in an all-to-all configuration, allows us to write the dynamical equation in terms of a global complex parameter, thus making explicit the mean field character of the model. Each oscillator appears to interact with the mean-field quantities $r$ and $\psi$ and uncoupled to each other. The phase of each oscillator $\theta_i$ is attracted to $\psi$ instead of another particular oscillator phase $\theta_j$. The strength of this interaction is governed by $r$, and as it increases, the collective behaviour strengthens and synchronization is achieved \cite{Strogatz00}. The model presented here has no way to define a global order parameter. Only locally can such variables be considered; therefore, there is no global mean variable to which the oscillator dynamics can be referred. Decoupling is impossible. This fact is essential to understand the results of the complexity picture described here. As reported in detail in \cite{garcia22}, when collective behaviour happens, there is not necessarily a global collective pack to which all oscillators are drawn. Instead, several local communities arise, and a non-trivial interaction between these communities can lead or not to global synchronization. The emergence of local competing communities lies at the heart of the appearance of complex behaviour. The local order parameters mean that each local community will have a value of $\psi_i$ different from the other communities. Therefore there is no unique phase value towards all oscillator units are attracted. The same idea also leads that there is no unique strength in the interactions, but instead a distribution $r_i$ of them. When analyzing  the time evolution of the complex local parameter for different $\omega, \gamma$, the steady-state value was found but not identical for each community. As reported here, these non-trivial collectively local dynamics give rise to complex spatiotemporal dynamics at certain values of the control parameters. Beyond the explanation of what causes the observed complexity, it is also clear that the model is capable of dynamics that can lead to enhanced computation capabilities.

\section{Conclusions}\label{Conclusions}

 A system of Adler-type oscillators, with Kuramoto-type interaction between nearest neighbours, is as simple as a model can be and yet, exhibits a wealth of behaviours, including complex regions with non-trivial regimes. This should not come entirely as a surprise, as similar features were found in a closely related system discussed previously by the authors. However, it does show that complexity does not stem from the particular choice of coupling. It is left to realize if the observed complexity is the result of the topology of the coupling and, therefore, which type of behaviours could be found in other couplings, such as the original all-to-all topology of Kuramoto, or small world topology, for example. 
 
 Compared to the previously studied Alonso model, this report's particular choice of coupling changes the control parameter diagrams, giving rise to two adjacent complex regions. Furthermore, a similar spike in $ECM$, together with jumps in $h$ and $d$, was found in a boundary region, but, in this case, and different from what was reported for CA and the Alonso coupled oscillators, here the boundary is between the absorbing and one complex region, not a chaotic boundary. However, it is unambiguous that the observed spike in $E$ is a signature of enhanced processing capabilities.    

 The spatiotemporal diagrams and the distance matrix analysis showed that the system could sustain non-trivial correlations.
 
The discretization of this type of system leads to a picture that resembles behaviours already seen in CA analysis which points to the degree of complexity that can be achieved. At the same time, in contrast with CA, we have a natural continuous change of the control parameters $\gamma$ and $\omega$. This means the oscillation system has a continuous space of ''rules'', where the parameters controlling these rules can take infinite values. The freedom that, in principle, this fact gives should call for better harvesting of the computational powers of the system compared to the unpredictability of the CA rule space. 

\section{Acknowledgment}\label{ack}

The University of Havana is acknowledged for financial support through an institutional project award. EER and KGM will like to thank Alexander von Humboldt Stiftung for financial support. EER and KGM will like to thank KIT for a great working environment.



\end{document}